\documentclass[twocolumn]{aastex62}

\newcommand{\fermi}{{\it Fermi}-LAT}
\newcommand{\gray}{$\gamma$-ray}
\newcommand{\grays}{$\gamma$-rays}
\usepackage{amsmath}

\received{-}
\revised{-}
\accepted{-}
\submitjournal{ApJ}

\shorttitle{Emission from CTA 102}
\shortauthors{Gasparyan et al.}

\begin{document}

\title{On the multi-wavelength Emission from CTA 102}

\correspondingauthor{N. Sahakyan}
\email{narek@icra.it}

\author{S. Gasparyan}
\affiliation{ICRANet-Armenia, Marshall Baghramian Avenue 24a, Yerevan 0019, Armenia.}

\author[0000-0003-2011-2731]{N. Sahakyan}
\affiliation{ICRANet-Armenia, Marshall Baghramian Avenue 24a, Yerevan 0019, Armenia.}
\affiliation{ICRANet, P.zza della Repubblica 10, 65122 Pescara, Italy.}

\author{V. Baghmanyan}
\affiliation{ICRANet-Armenia, Marshall Baghramian Avenue 24a, Yerevan 0019, Armenia.}
\affiliation{ICRANet, P.zza della Repubblica 10, 65122 Pescara, Italy.}

\author{D. Zargaryan}
\affiliation{ICRANet-Armenia, Marshall Baghramian Avenue 24a, Yerevan 0019, Armenia.}
\affiliation{ICRANet, P.zza della Repubblica 10, 65122 Pescara, Italy.}



\begin{abstract}
We report on broadband observations of CTA 102 ($z=1.037$) during the active states in 2016-2017. In the $\gamma$-ray band, Fermi LAT observed several prominent flares which followed a harder-when-brighter behavior: the hardest photon index $\Gamma=1.61\pm 0.10$ being unusual for FSRQs. The peak $\gamma$-ray flux above 100 MeV $(3.55\pm0.55)\times10^{-5}\:{\rm photon\:cm^{-2}\:s^{-1}}$ observed on MJD 57738.47 within 4.31 minutes, corresponds to an isotropic $\gamma$-ray luminosity of $L_{\gamma}=3.25\times10^{50}\:{\rm erg\:s^{-1}}$, comparable with the highest values observed from blazars so far. The analyses of the Swift UVOT/XRT data show an increase in the UV/optical and X-ray bands which is contemporaneous with the bright $\gamma$-ray periods. The X-ray spectrum observed by Swift XRT and NuSTAR during the $\gamma$-ray flaring period is characterized by a hard photon index of $\sim1.30$. The shortest e-folding time was $4.08\pm1.44$ hours, suggesting a very compact emission region $R\leq\delta\times2.16\times10^{14}$ cm. We modeled the spectral energy distribution of CTA 102 in several periods (having different properties in UV/optical, X-ray and $\gamma$-ray bands) assuming a compact blob inside and outside the BLR. We found that the high-energy data are better described when the infrared thermal radiation of the dusty torus is considered. In the flaring periods when the correlation between the $\gamma$-ray and UV/optical/X-ray bands is lacking, the $\gamma$-ray emission can be produced from the interaction of fresh electrons in a different blob, which does not make a dominant contribution at lower energies.
\end{abstract}



\section{Introduction}\label{sec:1}
The blazars are the most extreme class of radio-loud active galactic nuclei (AGNs) in their unification scheme. Blazars are emitting electromagnetic radiation ranging from radio to High and Very High Energy \gray{} bands (HE; $\geq100$ MeV and VHE; $\geq100$ GeV) characterized by rapid and high-amplitude variability which can be explained assuming the jets are oriented close to the line of sight of the observer (a few degrees) and the nonthermal plasma moves with relativistic velocities along the jet \citep{urry}. Blazars are grouped into two large sub-classes, Flat Spectrum Radio Quasars (FSRQs) and BL Lacertae objects (BL Lacs), on the basis of different emission line properties, which are stronger and quasar-like in FSRQs and weak or absent in BL Lacs. An alternative classification method is based on the luminosity of the broad emission lines (or accretion disc) measured in Eddington units: when $L_{\rm BLR}/L_{\rm Edd}\geq5\times10^{-4}$ the objects are FSRQs otherwise they are BL Lacs \citep{2011MNRAS,2012MNRAS}.\\
The multi-wavelength studies have shown that the Spectral Energy Distributions (SEDs) of both types of blazars consist of two broad humps, peaking in the IR-X-ray (low-energy component) and in the MeV-TeV bands (HE-component). The low-energy component is well explained by synchrotron emission from relativistic electrons in the jet, whereas the nature of the HE-component is less well understood as several different emission mechanisms can be responsible for that emission (e.g., see \cite{sikora09}). The simplest explanation scenario is the synchrotron self-Compton (SSC) radiation, where the soft synchrotron photons are inverse-Compton-up-scattered by the same electrons that have produced the synchrotron emission \citep{ghisellini, bloom, maraschi}. As FSRQ jets are in an environment with a stronger external radiation field which can be beamed and enhanced in the frame of the jet, the inverse Compton scattering of external photons too can contribute to the observed HE emission \citep{blazejowski,ghiselini09, sikora}. Alternatively, if the protons are efficiently accelerated in the jet (beyond the threshold for pion production), the HE emission can be also explained by the interaction of energetic protons \citep{mucke2,mucke1}.\\
After the lunch of Fermi Large Area Telescope (\fermi{}) several thousand blazars were detected in the \gray{} band \citep{ackermancatalog} which opens new perspectives for investigation of the broadband emission from them. The observations indirectly show that the \grays{} can be produced either close to or far from the central black hole. As the \gray{} emission regions are very compact, inferred from extreme short time scale variabilities (e.g., in minute scales \cite{2016ApJ,foschini11,foschini13, nalewajko, brown13, rani13,saito,hayashida15}) and that there is a sharp break in the GeV \gray{} spectra of some blazars \citep{,poutanen}, the emission is most likely produced within the broad-line regions (BLRs). On the other hand, the recent detection of $\geq100$ GeV photons from several FSRQs \citep{ahnen1441,aleksic1510, aleksic1222, 2017MNRAS.470.2861S} 
implies that the \gray{} emission region should most likely be beyond the BLR in order to bypass strong absorption of VHE photons \citep{poutanen, liu}. Unfortunately, the angular resolution of \gray{} instruments is not high enough (and will not be in the near future) to resolve and localize the \gray{} emission regions which makes it difficult to determine the exact origin of \gray{} emission from blazars as the jet dissipation can occur at any distance from the central black hole.\\ 
Among the FSRQs detected by \fermi{}, the powerful GeV \gray{} emitter CTA 102, $z=1.037$ \citep{schmidt}, is flaring frequently, its \gray{} flux sometimes exceeding  $10^{-5}\:{\rm photon\:s^{-1}\:cm^{-2}}$. CTA 102 is a luminous, well-studied highly polarized quasar \citep{stockman} having variable optical emission \citep{pica}. It has been initially identified by Compton Gamma Ray Observatory mission as a \gray{} emitter (the flux $>$ 100 MeV being $(2.4\pm0.5)\times10^{-7}\:{\rm photon\:s^{-1}\:cm^{-2}}$), and then it is being included in all the point source catalogs of \fermi{} \citep{acero}. Since 2016, CTA 102 was in the enhanced emission state in the UV/optical, X-ray and HE \gray{} bands \citep{casadio, Balonek, chapman, popov, ciprini, bulgarelli, ciprini1, becerra, minervini, carrasco} with several prominent \gray{} bright periods. Considering the available large amount of multi-wavelength data which allows to constrain the emitting region size and location, magnetic field and electron energy distribution, etc., CTA 102 is an ideal object for exploring the physics of FSRQ jets.\\
In this paper, we analyze the Swift UVOT/XRT, NuSTAR and \fermi{} data collected from 2016 to 2018 to study the broadband emission from CTA 102. The data collected for the analysis and its reduction methods are described in Section \ref{sec:2}. The spectral changes in different bands during the flaring and low state is discussed in Section \ref{sec:3}. The broadband SED modeling is presented in Section \ref{sec:4} and Results and Discussion in Section \ref{sec:5}. The conclusion is summarized in Section \ref{sec:6}.
\section{Observations and Data Reduction}\label{sec:2}
\subsection{Gamma-ray observations: Fermi LAT}
\begin{figure*}
  \centering
   \includegraphics[width=0.99 \textwidth]{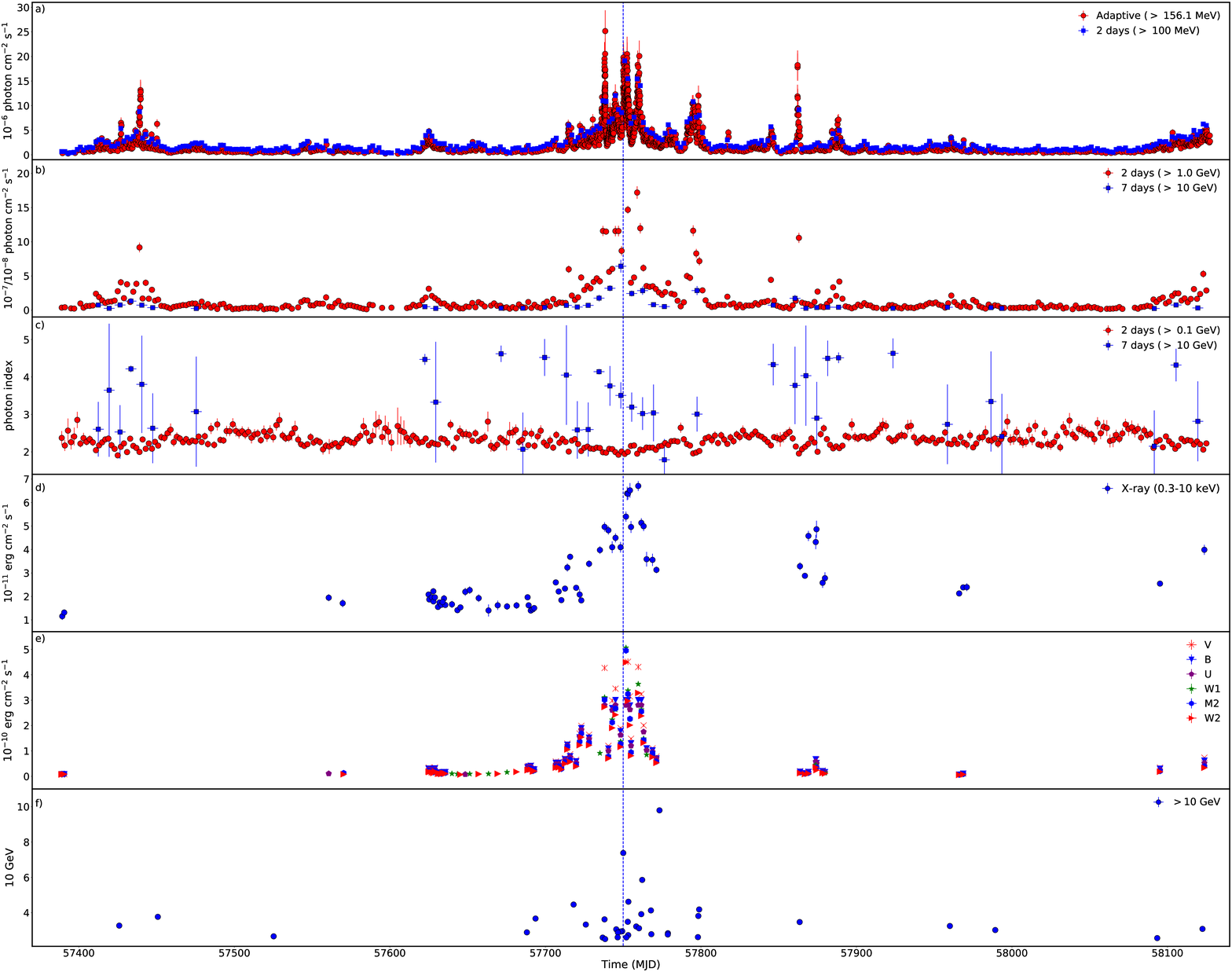}
    \caption{Multifrequency light curve of CTA 102 obtained for the period from 2008 August to 2018 January. {\it a)} \gray{} light curves with adaptive (red; $\geq156.1$ MeV) and 2-day (blue; $100$ MeV) bins, {\it b) and c)} the flux and photon index with 2- and 7-days binning, {\it d)} Swift XRT light curve in the 0.3-10 keV range, {\it e)} UV/optical fluxes in $V$, $B$, $U$, $W1$, $M2$ and $W2$ bands and {\it f)} the energy and arrival times of the highest-energy photons. The vertical blue dashed line shows the period when a large flare in the $R-$ band was observed (28 December 2016).
   }%
    \label{var_mult}
\end{figure*}
In the present paper we use the publicly available \fermi{} data acquired in the period from 01 January 2016 to 09 January 2018 when large-amplitude flares of CTA 102 were observed. Fermi Science Tools v10r0p5 was used to analyze the data with {\it P8R2\_SOURCE\_V6} instrument response function. Only the 100 MeV - 300 GeV events extracted from a $12^\circ$ region of interest (ROI) centered on the location of CTA 102 [(RA,dec)= (338.139, 11.720)] have been analyzed. However, the results were checked by repeating the same analyses selecting ROI radii of $10^\circ$ and $15^\circ$. To eliminate the Earth limb events, the recommended quality cuts, (DATA\_QUAL==1)$\&\&$(LAT\_CONFIG==1) and a zenith angle cut at $90^\circ$ were applied. After binning the data into pixels of $0.1^\circ\times0.1^\circ$ and into 34 equal logarithmically-spaced energy bins, with the help of {\it gtlike} a binned likelihood analysis is performed. 
The model file describing ROI was created using the \fermi{} third source catalog \citep{acero} (3FGL) which contains sources within ROI+$5^\circ$ from the target, as well as Galactic {\it gll\_iem\_v06} and {\it iso\_P8R2\_SOURCE\_V6\_v06} diffuse components. All point-source spectra were modeled with those given in the catalog, allowing the photon index and normalization of the sources within $12^\circ$ to be free in the analysis. Also, the normalization of diffuse background components are free. To check if there are new \gray{} sources in the ROI,  a Test Statistics (TS) map (TS defined as TS $= 2(log {\rm L}-log {\rm L_0})$, where ${\rm L}$ and ${\rm L_0}$ are the likelihoods whether or not the source is included) is created with {\it gttsmap} tool which places a point source at each pixel and evaluates its TS. In the TS map, there are new hotspots (pixels) with TS $>$ 25 ($5\:\sigma$) which possibly hints at the presence of new sources. For each new hotspot we sequentially added a new point source with a power-law spectral definition. For the further analysis the model file containing these additional point sources is used.\\
In the whole-time analysis, the \gray{} spectrum of CTA 102 was first modeled using a log-parabola \citep{Massaro} as in 3FGL and then assuming a power-law shape. The latter will be used in the light curve calculations, as shorter periods will be considered and a power law can be a good approximation of the spectrum. During the analysis of each individual flare a different model file obtained from the analyses of the data accumulated during one/two- month periods covering the flares was also used. An unbinned maximum likelihood analysis was performed using $(0.1-300)$ GeV photons with the appropriate quality cuts mentioned above, to obtain the \gray{} light curves. Since no variability is expected from the underlying background diffuse emission, we fix the normalization of both background components to the best fit values obtained for the whole time period.\\
Initially, the light curve was calculated with the help of an adaptive binning method. At regular (fixed) time binning, the long bins will smooth out the fast variation while short bins might result in many upper limits during the low-activity periods. In the adaptive binning method, the time bin widths are adjusted to produce bins with constant flux uncertainty above the optimal energies \citep{lott} meant to find rapid changes in \gray{} fluxes. The adaptively binned light curve with 15\% uncertainty and above $E_0=156.1$ MeV in Fig. \ref{var_mult} shows several bright \gray{} states: from MJD 57420 to MJD 57445 and from  MJD 57700 to MJD 57900. The peak flux of $(2.52\pm0.42)\times10^{-5}\:{\rm photon\:cm^{-2}\:s^{-1}}$ with a photon index of $\Gamma=1.99\pm0.15$ was observed on MJD 57738.47 within 4.31 minutes with a convincingly high $\sim20.0\sigma$. It corresponds to a flux of $(3.55\pm0.55)\times10^{-5}\:{\rm photon\:cm^{-2}\:s^{-1}}$ above 100 MeV which $\sim221$ times exceeds the average \gray{} flux given in 3FGL ($\simeq1.60\times10^{-7}\:{\rm photon\:cm^{-2}\:s^{-1}}$ but the source is variable with a variability index of 1602.3 in 3FGL). In addition, we used {\it gtfindsrc} tool to determine the best coordinates of the \gray{} emission in this period, yielding (RA,dec)= (338.115, 11.746) with a 95\% confidence error circle radius of $r_{95}$ = 0.06. These coordinates are offset only by $0.03^\circ$ from the \gray{} position of CTA 102, indicating that it is the most likely source of the emission. The hardest photon index of $1.61\pm 0.10$ ($22.56\sigma$) was observed on MJD 57752.45 within $9.46$ minutes, which is significantly harder than the mean photon index observed during the considered period, $\Gamma_{\rm mean}=2.22$. \\
In the adaptively binned light curve there is a hint at flux changes in minute scales. For example, the interval of MJD 57737.88- MJD 57739.00 ($\sim1.13$ days), contains 67 adaptive bins each having a width of the order of a few minutes and a detection significance of $> 14.3\sigma$. Another such active period was observed on MJD 57752.0, though the time bin widths were a few tens of minute. Many times during the considered period, the source flux exceeded $10^{-5}\:{\rm photon\:cm^{-2}\:s^{-1}}$, mostly observed during the extremely active period from MJD $57736.4$ to MJD $57798.46$ as well as a few times on MJD 57439.0 and MJD 57862.0. During these periods, the photon flux and index vary within $(1.01-2.52)\times10^{-5}\:{\rm photon\:cm^{-2}\:s^{-1}}$ and $1.61-2.56$, respectively, the minimum and maximum bin widths being $4.31$ and $194.54$ minutes and the detection significance varying from $13.18\sigma$ to $22.61\sigma$.\\
\begin{figure}
\centering
\includegraphics[width=\hsize,clip]{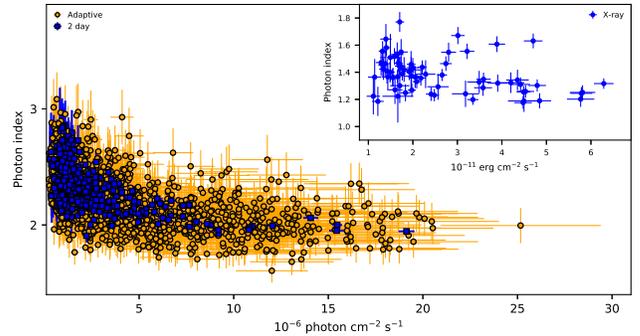}
\caption{CTA 102 \gray{} photon index vs. flux in adaptive (orange) and two-day bins (blue). Similar plot for the X-ray band is shown in the insert.}
\label{phot_index}
\end{figure}
Fig. \ref{var_mult} b) shows the \gray{} light curve $>1$ GeV (2 days; red color) and $>10$ GeV (7 days; blue color) with a noticeable increase in the flux, the peaks being
$(2.32\pm0.10)\times10^{-6}\:{\rm photon\:cm^{-2}\:s^{-1}}$ and $(6.43\pm0.94)\times10^{-8}\:{\rm photon\:cm^{-2}\:s^{-1}}$, at 2-day and 7-day binning, respectively. Above 10 GeV, among 105 total bins only in 36 the detection significance is at least $4\sigma$, but, e.g., on MJD 57741.0 and MJD 57748.0 it is as large as $\simeq29\sigma$, $N_{\rm pred}$ varying within $46-55$.
The \gray{} photon index variation above 0.1 and 10 GeV is shown in Fig. \ref{var_mult} c) with red and blue colors, respectively. There is an obvious hardening above 0.1 GeV, when the photon index changed to $\Gamma\simeq2.0$, during the most bright periods of the source. The mean \gray{} photon index above 10 GeV is $\Gamma_{\rm mean}=3.41$ but on MJD 57776.0 $\Gamma=1.79\pm0.55$ with $7.85\sigma$.\\
The \gray{} photon index versus flux is presented in Fig. \ref{phot_index} for adaptive (orange) and 2-day binning (blue; $> 0.1$ GeV). When 2-day intervals are considered, there is a hint of spectral hardening as the source gets brighter. In the \gray{} band such behaviour has been already observed from several blazars (e.g., PKS 1502+106 \citep{2010ApJ...710..810A}, PKS 1510-089 \citep{2010ApJ...721.1425A}, sometimes from 3C 454.3 \citep{2010ApJ...721.1383A}, etc.) and radio galaxies (e.g., NGC 1275 \citep{2017ApJ...848..111B}). Such evolution of spectral index and flux is expectable when accelerated HE electrons are cooled down (e.g., \citep{kirk}). It is hard to see similar relation in the case of adaptive bins as the bright periods last shorter, leading to larger uncertainties. The linear-Pearson correlation test applied to 2-day and adaptively binned intervals yielded $r_{p}=-0.569$ and $r_{p}=-0.533$, respectively, the p-value being $<<10^{-5}$. This suggests negative correlation between the flux and photon index, i.e., as the flux increases, the photon index decreases (hardens).\\
The distribution of highest energy events ($>10$ GeV) detected from CTA 102, calculated using the {\it gtsrcprob} tool is presented in Fig. \ref{var_mult} f). Most of the HE photons are observed during MJD 57700-57800 with the maximum of $97.93$ GeV detected on MJD 57773.34.
\subsection{Swift UVOT/XRT observations}
The data from seventy Swift (Neil Gehrels Swift observatory) observations of CTA 102 carried out from 01 January 2016 to 09 January 2018 have been analyzed. The exposures range from 0.3 ks (ObsID:33509083) to 3.14 ks (ObsID:33509095) and most of the observations were made in the photon counting and only two in the window timing mode. The XRT data were analyzed with {\it XRTDAS} (v.3.3.0) using standard procedure and the most recent calibration databases. Events for the spectral analysis were selected within a 20 pixel ($47''$) circle with the source at the center, while the background region as an annulus with the same center and having inner and outer radii of 51 ($120''$) and 85 pixels ($200''$), respectively. The count rate in some observations was above 0.5 count ${\rm s^{-1}}$ implying pile-up in the inner part of the PSF. This effect was removed by excluding the events within a 3 pixel radius circle centered on the source position. The Cash statistics \citep{1979ApJ...228..939C} on ungrouped data was used as for some observations the number of counts was low. However for the observations with a high count rate, the results  were also cross-checked by rebining to have at least 20 counts per bin and then fitted using the $\chi^2$ minimization technique. The individual spectra were fitted with {\it XSPEC} v12.9.1a adopting an absorbed power-law model with $N_{H}=5.35\times 10^{20}\:{\rm cm^{-2}}$ column density, ignoring the channels with energy below 0.3 keV and above 10 keV. Fig. \ref{var_mult} d) shows the X-ray flux evolution in time (the corresponding parameters are presented in Table \ref{tabeswift}), where its gradual increase contemporaneous with the \gray{} flux around MJD 57750 can be seen. The highest flux of $F_{0.3-10\:{\rm keV}}\simeq(6.71\pm 0.21)\times10^{-11}\:{\rm erg\:cm^{-2}\:s^{-1}}$ observed on MJD 57759.69 exceeds the average flux ($\simeq1.2\times10^{-11}\:{\rm erg\:cm^{-2}\:s^{-1}}$) $\sim5.6$ times. A relation between the unabsorbed X-ray flux and photon index is represented in the insert of Fig. \ref{phot_index}. A trend of a harder spectrum when the source is brighter can be seen. Such harder-when-brighter trend in the X-ray band was already observed from several FSRQs (e.g., PKS 1510-089 \citep{2008ApJ...672..787K, 2011A&A...529A.145D}, 3C 454.3 \citep{2010ApJ...712..405V} and etc.) which can be described if assuming the electrons are losing energy mainly through interaction with the external photon fields (e.g., \citep{2011ApJ...736L..38V}).\\
The data from the second instrument on board the Swift satellite, UVOT, was used to measure the flux of the source in the UV/optical bands. Photometry was computed using a five-arcsecond source region around CTA 102 and for the background - a source-free annulus centered on the source position with $27''$ inner and $35''$ outer radii. The magnitudes were computed using {\it UVOTSOURCE} task, then corrected for extinction, using the reddening coefficient E(B-V) from \citet{schlafly} and the ratios of the extinction to reddening A/E(B-V) for each filter from \citet{fitzpatrick} then converting to fluxes, following \citet{breeveld}. The flux measured for $V,\:B, \: U, \: W1, \:M2$ and $W2$ filters is shown in Fig. \ref{var_mult} e). Even if the available data are not enough for detailed studies, it is clear that up to $\sim$ MJD 57720 the source was in a relatively faint state in the optical/UV band but its flux significantly increased during the bright flaring period around $\sim$ MJD 57750. This is in agreement with the recent results by \citet{2017Natur.552..374R} which show that the source emission in the optical band increased in late 2016 with a  6-7 magnitude jump as compared with the minimal state. The maximum flux in the $R-$ band was observed on 28 December 2016 (MJD 57750) with a peak luminosity of  $1.32\times10^{48}\:{\rm erg\:s^{-1}}$. In addition, the radio monitoring (at 37 GHz) showed that the peak in this band is much earlier than the one in the R-band, inferring these emissions were produced in different locations of the jet.
\subsection{NuSTAR observation}
\begin{figure}
\centering
\includegraphics[width=\hsize,clip]{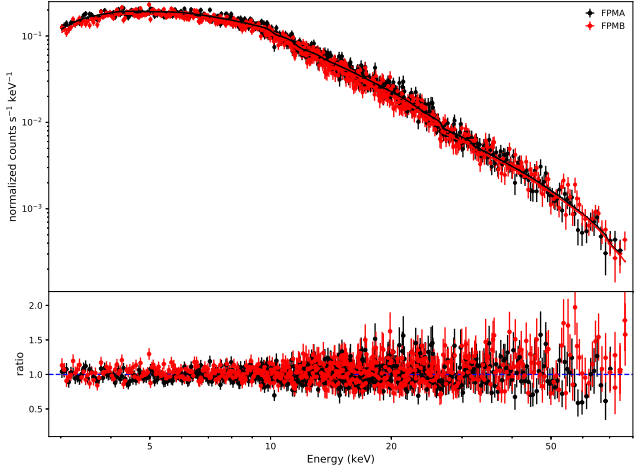}
\caption{{\it Top:} NuSTAR FPMA (black) and FPMB (red) spectra and best-fit models. {\it Bottom:} Residuals with respect to power-law model.}
\label{Nustarspec}
\end{figure}

In the hard X-ray band (3-79 keV), CTA 102 was observed once on 30 December 2016 by NuSTAR with a net exposure of $\sim26.21$ ks, when it was bright in the X-ray and \gray{} bands. The raw data (from both Focal Plane Modules [FPMA and FPMB; \citep{2013ApJ...770..103H}] were processed with the NuSTAR Data Analysis Software ({\it NuSTARDAS}) package v.1.4.1 (via the script {\it nupipeline}), producing calibrated and cleaned event files. The events data were extracted from a region of  $75''$ centered on the source position, while the background was extracted from a nearby source free circular region with the same radius. The spectra were binned so to have at least 30 counts per bin and fitted assuming an absorbed power-law model. The best fit resulted in $\Gamma_{\rm X}=1.32\pm0.005$ and $F_{3-79\:{\rm keV}}\simeq(2.94\pm 0.02)\times10^{-10}\:{\rm erg\:cm^{-2}\:s^{-1}}$ with $\chi^2=0.97$ for 1131 degrees of freedom. The corresponding {spectra for FPMA and FPMB} are shown in Fig. \ref{Nustarspec}.
\startlongtable
\begin{deluxetable*}{cccccc}
\tablecaption{Summary of Swift XRT observations of CTA 102. \label{tabeswift}}
\tablehead{
\colhead{Sequence No.}  & \colhead{Date (MJD)} &  \colhead{Exp(sec)} & \colhead{Log(Flux)\tablenotemark{a}} & $\Gamma$ & \colhead{C-stat./dof}
}
\tabletypesize{\footnotesize}
\startdata
00033509016 & 2016-01-02(57389.33) & 834.1 & $-10.94\pm0.06$ & $1.23\pm0.14$ & 91.23(103) \\
00033509017 & 2016-01-03(57390.6) & 1119.0 & $-10.88\pm0.04$ & $1.18\pm0.10$ & 166.99(168) \\
00033509018 & 2016-06-21(57560.67) & 991.4 & $-10.71\pm0.03$ & $1.43\pm0.08$ & 218.22(218) \\
00033509021 & 2016-06-30(57569.66) & 844.1 & $-10.77\pm0.04$ & $1.42\pm0.11$ & 195.86(151) \\
00033509022 & 2016-08-24(57624.88) & 1633.0 & $-10.68\pm0.02$ & $1.38\pm0.06$ & 422.02(320) \\
00033509023 & 2016-08-25(57625.35) & 1691.0 & $-10.73\pm0.02$ & $1.46\pm0.06$ & 393.49(306) \\
00033509024 & 2016-08-26(57626.94) & 1868.0 & $-10.71\pm0.02$ & $1.41\pm0.06$ & 410.73(323) \\
00033509025 & 2016-08-27(57627.94) & 1466.0 & $-10.65\pm0.03$ & $1.43\pm0.07$ & 233.39(291) \\
00033509026 & 2016-08-28(57628.01) & 2148.0 & $-10.75\pm0.02$ & $1.45\pm0.06$ & 271.35(309) \\
00033509027 & 2016-08-28(57628.94) & 2797.0 & $-10.71\pm0.02$ & $1.36\pm0.05$ & 450.67(403) \\
00033509028 & 2016-08-30(57630.93) & 1576.0 & $-10.81\pm0.03$ & $1.48\pm0.07$ & 226.96(270) \\
00033509030 & 2016-08-31(57631.93) & 2133.0 & $-10.76\pm0.03$ & $1.35\pm0.07$ & 289.41(290) \\
00033509031 & 2016-09-02(57633.06) & 1978.0 & $-10.78\pm0.02$ & $1.42\pm0.06$ & 316.53(301) \\
00033509034 & 2016-09-03(57634.79) & 966.5 & $-10.72\pm0.03$ & $1.57\pm0.09$ & 220.83(194) \\
00033509035 & 2016-09-04(57635.64) & 869.1 & $-10.79\pm0.03$ & $1.64\pm0.09$ & 207.90(193) \\
00033509076 & 2016-09-02(57633.92) & 1965.0 & $-10.75\pm0.02$ & $1.47\pm0.06$ & 373.93(324) \\
00033509077 & 2016-09-08(57639.9) & 991.4 & $-10.78\pm0.04$ & $1.33\pm0.10$ & 202.35(178) \\
00033509078 & 2016-09-12(57643.43) & 914.0 & $-10.85\pm0.04$ & $1.47\pm0.10$ & 160.64(171) \\
00033509079 & 2016-09-14(57645.36) & 1091.0 & $-10.81\pm0.03$ & $1.4\pm0.09$ & 192.03(199) \\
00033509080 & 2016-09-17(57648.47) & 894.0 & $-10.66\pm0.03$ & $1.34\pm0.08$ & 262.16(217) \\
00033509081 & 2016-09-20(57651.32) & 996.4 & $-10.64\pm0.03$ & $1.33\pm0.07$ & 311.96(242) \\
00033509082 & 2016-09-26(57657.11) & 789.1 & $-10.72\pm0.04$ & $1.43\pm0.09$ & 198.62(189) \\
00033509083 & 2016-10-02(57663.43) & 344.6 & $-10.85\pm0.07$ & $1.37\pm0.20$ & 47.08(67) \\
00033509084 & 2016-10-08(57669.34) & 609.3 & $-10.79\pm0.05$ & $1.45\pm0.12$ & 130.49(103) \\
00033509085 & 2016-10-14(57675.33) & 966.5 & $-10.8\pm0.04$ & $1.38\pm0.10$ & 221.53(186) \\
00033509086 & 2016-10-20(57681.43) & 971.4 & $-10.79\pm0.04$ & $1.32\pm0.09$ & 248.18(190) \\
00033509087 & 2016-10-27(57688.54) & 1965.0 & $-10.71\pm0.02$ & $1.29\pm0.06$ & 449.38(329) \\
00033509088 & 2016-10-28(57689.21) & 1711.0 & $-10.79\pm0.03$ & $1.36\pm0.07$ & 301.62(271) \\
00033509090 & 2016-10-29(57690.59) & 1723.0 & $-10.85\pm0.03$ & $1.46\pm0.08$ & 182.50(224) \\
00033509091 & 2016-10-30(57691.92) & 1656.0 & $-10.84\pm0.03$ & $1.4\pm0.08$ & 231.46(241) \\
00033509092 & 2016-10-31(57692.79) & 2108.0 & $-10.82\pm0.02$ & $1.57\pm0.06$ & 287.31(299) \\
00033509093 & 2016-11-14(57706.68) & 2974.0 & $-10.59\pm0.02$ & $1.22\pm0.04$ & 597.15(447) \\
00033509094 & 2016-11-16(57708.53) & 2762.0 & $-10.66\pm0.02$ & $1.25\pm0.05$ & 460.61(428) \\
00033509095 & 2016-11-18(57710.26) & 3137.0 & $-10.73\pm0.02$ & $1.33\pm0.05$ & 432.80(415) \\
00033509096 & 2016-11-20(57712.58) & 2435.0 & $-10.63\pm0.02$ & $1.32\pm0.05$ & 556.82(417) \\
00033509097 & 2016-11-22(57714.11) & 1693.0 & $-10.49\pm0.02$ & $1.55\pm0.06$ & 265.95(322) \\
00033509098 & 2016-11-23(57715.9) & 2934.0 & $-10.43\pm0.02$ & $1.19\pm0.04$ & 717.13(517) \\
00033509099 & 2016-11-27(57719.78) & 1963.0 & $-10.63\pm0.02$ & $1.36\pm0.05$ & 505.78(364) \\
00033509100 & 2016-11-30(57722.02) & 382.1 & $-10.68\pm0.05$ & $1.42\pm0.12$ & 108.37(118) \\
00033509101 & 2016-12-01(57723.08) & 1341.0 & $-10.74\pm0.02$ & $1.78\pm0.07$ & 278.11(275) \\
00033509103 & 2016-12-06(57728.07) & 1958.0 & $-10.47\pm0.02$ & $1.69\pm0.05$ & 449.08(354) \\
00033509105 & 2016-12-13(57735.06) & 2655.0 & $-10.40\pm0.02$ & $1.32\pm0.04$ & 457.45(437) \\
00033509106 & 2016-12-16(57738.05) & 2440.0 & $-10.30\pm0.02$ & $1.23\pm0.04$ & 653.22(469) \\
00033509107 & 2016-12-18(57740.49) & 2402.0 & $-10.32\pm0.02$ & $1.27\pm0.05$ & 569.51(541) \\
00033509108 & 2016-12-20(57742.95) & 818.4 & $-10.39\pm0.03$ & $1.47\pm0.08$ & 271.95(359) \\
00033509109 & 2016-12-23(57745.07) & 1993.0 & $-10.35\pm0.02$ & $1.58\pm0.05$ & 399.45(388) \\
00033509110 & 2016-12-26(57748.33) & 1686.0 & $-10.39\pm0.02$ & $1.39\pm0.06$ & 347.1(329) \\
00033509111 & 2016-12-29(57751.8) & 1823.0 & $-10.27\pm0.02$ & $1.62\pm0.04$ & 455.60(397) \\
00033509112 & 2016-12-30(57752.54) & 1468.0 & $-10.19\pm0.02$ & $1.29\pm0.05$ & 482.04(410) \\
00088026001 & 2016-12-31(57753.06) & 2048.0 & $-10.20\pm0.02$ & $1.26\pm0.04$ & 619.50(486) \\
00033509113 & 2017-01-02(57755.05) & 1566.0 & $-10.30\pm0.02$ & $1.24\pm0.05$ & 383.85(386) \\
00033509114 & 2017-01-01(57754.37) & 1488.0 & $-10.19\pm0.02$ & $1.18\pm0.05$ & 405.54(421) \\
00033509115 & 2017-01-06(57759.69) & 2472.0 & $-10.17\pm0.01$ & $1.33\pm0.03$ & 748.83(539) \\
00033509116 & 2017-01-08(57761.68) & 2480.0 & $-10.29\pm0.02$ & $1.31\pm0.04$ & 507.41(465) \\
00033509117 & 2017-01-10(57763.14) & 2502.0 & $-10.30\pm0.02$ & $1.17\pm0.04$ & 607.92(463) \\
00033509118 & 2017-01-12(57765.07) & 521.9 & $-10.45\pm0.04$ & $1.19\pm0.09$ & 200.60(200) \\
00033509119 & 2017-01-15(57768.86) & 1009.0 & $-10.45\pm0.03$ & $1.33\pm0.08$ & 254.30(243) \\
00033509120 & 2017-01-18(57771.38) & 1768.0 & $-10.50\pm0.02$ & $1.41\pm0.05$ & 399.09(391) \\
00033509121 & 2017-04-20(57863.68) & 1975.0 & $-10.48\pm0.02$ & $1.56\pm0.06$ & 342.39(331) \\
00033509122 & 2017-04-23(57866.86) & 2273.0 & $-10.54\pm0.02$ & $1.38\pm0.05$ & 467.86(419) \\
00033509123 & 2017-04-26(57869.13) & 2018.0 & $-10.34\pm0.02$ & $1.33\pm0.05$ & 494.03(383) \\
00033509124 & 2017-04-30(57873.83) & 991.4 & $-10.36\pm0.03$ & $1.34\pm0.07$ & 298.45(263) \\
00033509125 & 2017-05-01(57874.31) & 891.5 & $-10.31\pm0.03$ & $1.16\pm0.08$ & 207.36(245) \\
00033509126 & 2017-05-05(57878.23) & 681.8 & $-10.59\pm0.04$ & $1.41\pm0.09$ & 203.60(192) \\
00033509127 & 2017-05-06(57879.75) & 529.4 & $-10.56\pm0.04$ & $1.33\pm0.09$ & 205.28(182) \\
00033509128 & 2017-08-01(57966.04) & 1975.0 & $-10.67\pm0.02$ & $1.45\pm0.05$ & 427.29(342) \\
00033509129 & 2017-08-03(57968.65) & 2298.0 & $-10.62\pm0.02$ & $1.42\pm0.05$ & 457.95(394) \\
00033509131 & 2018-01-05(58123.69) & 1970.0 & $-10.4\pm0.02$ & $1.27\pm0.06$ & 397.65(354) \\
00033509130 & 2017-08-05(57970.96) & 876.5 & $-10.62\pm0.03$ & $1.41\pm0.08$ & 270.04(241) \\
00033509061 & 2017-12-08(58095.17) & 2477.0 & $-10.59\pm0.02$ & $1.25\pm0.05$ & 444.7(416) \\ 
\hline
\enddata
\tablenotetext{a} {Flux in 0.3--10 keV in unit of erg cm$^{-2}$ s$^{-1}$}.
\end{deluxetable*}

\subsection{The light curves variability}
\begin{figure}
  \centering
    \includegraphics[width= 0.475 \textwidth]{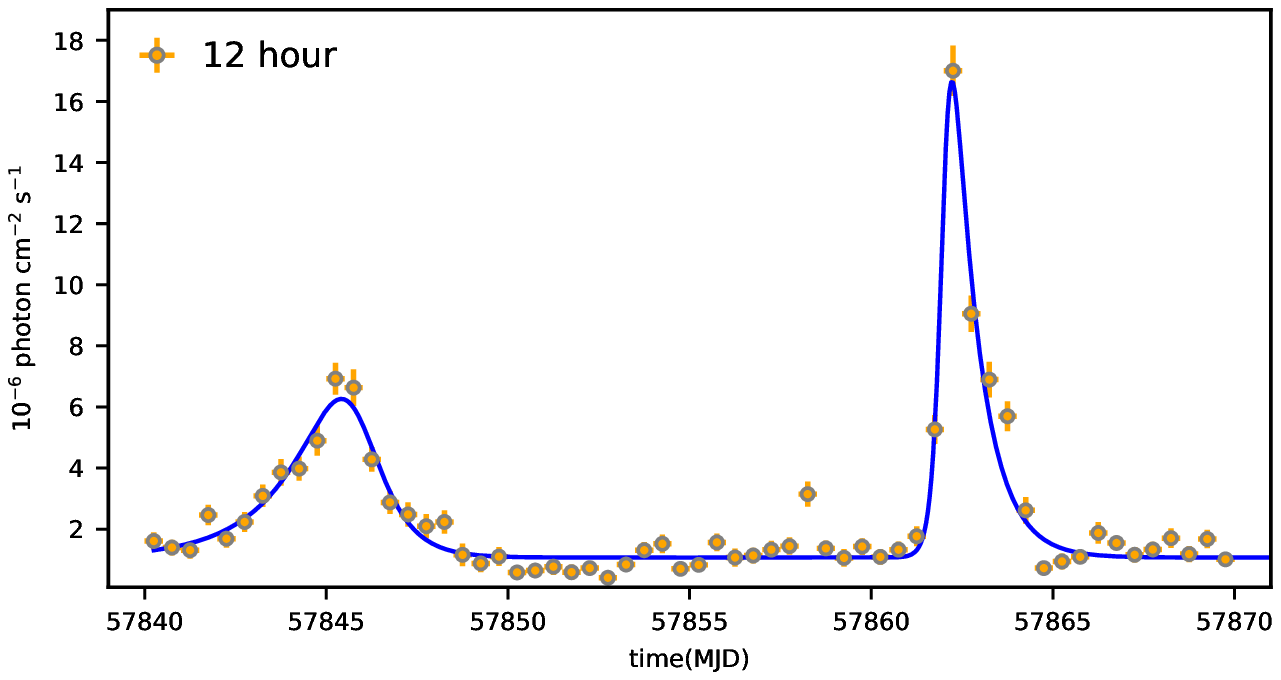}
    \includegraphics[width= 0.475 \textwidth]{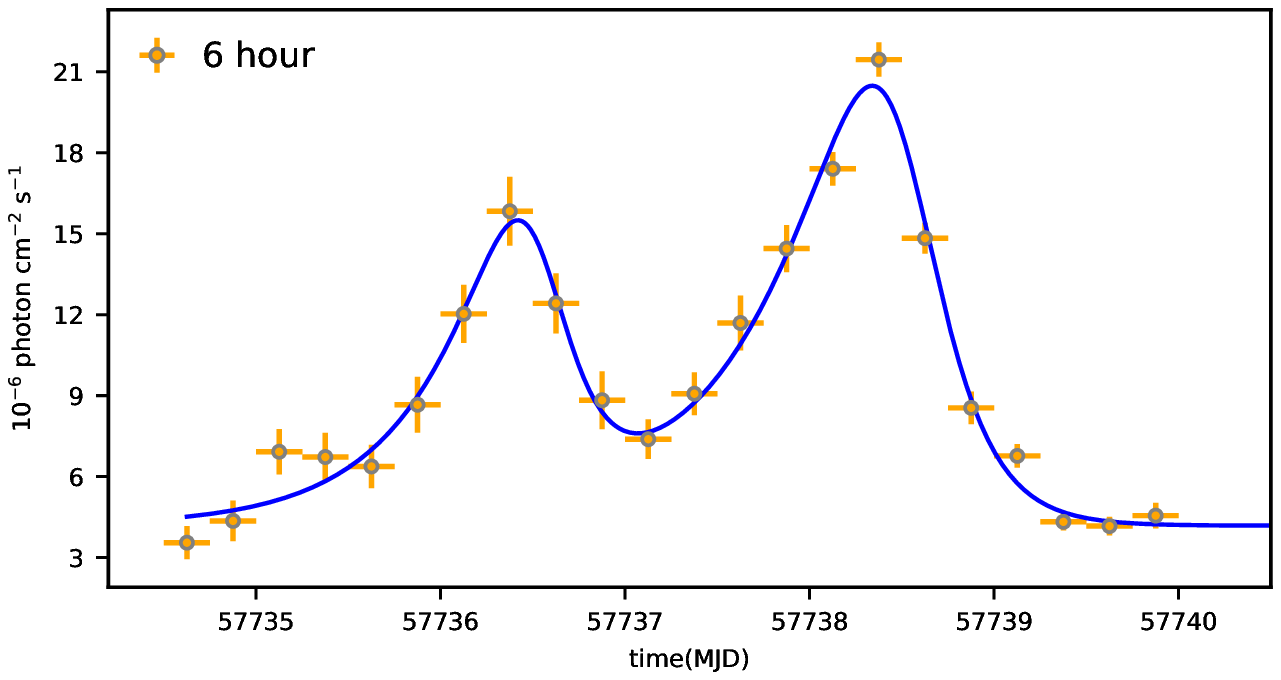}
    \caption{Light curves of CTA 102 above 100 MeV with time binning of 6 h (upper panel) and 12 h (lower panel). The lines show the flare fit with Eq. \ref{func1} (Table \ref{fit_par}).
   }%
    \label{flares}
\end{figure}
\begin{deluxetable}{cccccc}
\tablecaption{Parameter values best explaining the flares. \label{fit_par}}
\tablehead{
\colhead{Flare period $t_{0}$}  & \colhead{$t_r \pm err$} & \colhead{$t_d \pm err$} & $F_{\rm 0}/10^{-6}$  \\
\colhead{MJD} & \colhead{(day) } & \colhead{(day)} & \colhead{$\mathrm{\rm photon\:cm}^{-2}s^{-1}$}
}
\startdata
$57736.53\pm 0.11$\tablenotemark{a} &  $0.46 \pm 0.13$ & $0.17 \pm 0.08$  & $18.68 \pm 3.33$ \\
$57738.50\pm 0.06$\tablenotemark{a} &  $0.60 \pm 0.09$ & $0.21\pm 0.03$  & $29.04\pm2.39$  \\ 
$57845.78\pm 0.36$\tablenotemark{b} &  $1.49\pm0.33$ & $0.70\pm0.23$   & $9.72\pm 1.26$  \\
$57862.02\pm 0.11$\tablenotemark{b} &  $0.17\pm0.06$ & $0.73\pm0.11$  & $25.20\pm2.63$  \\
\enddata
 \tablenotetext{a}{$F_{\rm c}=(4.18\pm 0.34)\times 10^{-6}\mathrm{\rm photon\:cm}^{-2}s^{-1}$.}
 \tablenotetext{b}{$F_{\rm c}=(1.07\pm 0.08)\times 10^{-6}\mathrm{\rm photon\:cm}^{-2}s^{-1}$.}
\end{deluxetable}
The \gray{} (2-day ($>0.1$ and $>1.0$ GeV), 7-day ($>10.0$ GeV) and adaptive binned ($>156.1$ MeV)), X-ray (0.3-10 keV) and UV/optical fluxes variation in time are shown in the a), b), c), d) and e) panels of Fig. \ref{var_mult}. There is an evident major \gray{} flux increase accompanied by moderate brightening in the X-ray and UV/optical bands. The variability in different bands is quantified using their fractional rms variability ($F_{\rm var}$) amplitude \citep{vaughan}, resulting in $F_{\rm var}=0.511\pm 0.008$ for X-ray band and correspondingly $0.920\pm 0.006$ and $0.984\pm 0.004$  for the \gray{} light curves with adaptive and 2-day ($>0.1$ GeV) binning, implying much stronger variability in the \gray{} band. This variability is even stronger when the light curves with 2-day ($>1.0$ GeV) and 7-day ($10.0$ GeV) bins are used (excluding correspondingly 20 and 69 periods with upper limits in them), since $F_{\rm var}=1.61\pm 0.01$ and $1.18\pm 0.06$, respectively.\\
The rapid variability in the \gray{} band can be further investigated by fitting the data with the double exponential form function to obtain the time profiles of the flux variations.  However we note that the double exponential form function is not unique and the flare time profiles can be reproduced also by other functions (e.g., see \citet{2010ApJ...714L..73A}). As the main purpose of the current fit is only to estimate the rise and decay times, we fit the light curves with the following function \citep{abdoflares}:
\begin{equation}
F(t)= F_{\rm c}+F_0\times\left(e^{\frac{t-t_{\rm 0}}{t_{\rm r}}}+e^{\frac{t_{\rm 0}-t}{t_{\rm d}}}\right)^{-1}
\label{func1}
\end{equation}
where $t_0$ is the time of the flare peak ($F_0$) and $t_r$ and $t_d$ are the rise and decay times, respectively. Each light curve was fitted with the non-linear optimization python package {\it lmfit} \footnote{https://lmfit.github.io/lmfit-py/} using a function that contains two inverses of the sum of exponentials (corresponding to the number of flares).\\   
The active (bright) periods identified in the adaptively binned light curve are analyzed with normal time sampling and only the periods when the rise and decay times can be well constrained are considered. Accordingly, the periods from MJD 57734 to MJD 57740 and from MJD 57840 to MJD 57870 (Fig. \ref{flares}) divided into 6- and 12 hour bins respectively are selected; the detection significance in each bin is $>5\sigma$ and the plot of Npred/$\sqrt{\rm Npred}$ vs Flux/$\Delta {\rm Flux}$ shows linear correlation, so the likelihood fit converged for each time bins. The identified four peaks are sequentially numbered from 1 to 4 (F1- F4).\\
The fit is shown in Fig. \ref{flares} and the corresponding parameters are given in Table \ref{fit_par}. The average flux level ($F_{\rm c}$) is left free during the fitting and the corresponding values are presented in Table \ref{fit_par}. The flares 1-3 have rise times longer than the fall, and only F4 shows the opposite tendency. The symmetry of the flares can be quantitatively estimated by calculating the parameter of $\xi=(t_{\rm d}-t_{\rm r})/(t_{\rm d}+t_{\rm r})$ as defined in \citet{abdoflares} which ranges from $-0.64$ to $-0.46$ for F1-3 and $0.62$ for F4, implying these are moderately asymmetric flares. The shortest e-folding times for rise and decay are $t_{\rm r}=0.17\pm0.06$ and $t_{\rm d}=0.21\pm 0.03$ day  \footnote{in Table \ref{fit_par} e-folding times are given, the doubling or halving timescales can be computed by $t_{\rm r,d} \times ln2$} observed during F2 and F4, respectively. During F4, when the highest flux was observed within $4.08\pm1.44$ hours, the flux increased up to $(2.52\pm0.26)\times10^{-5}\mathrm{\rm photon\:cm}^{-2}s^{-1}$ and dropped to its average level within $17.52\pm2.64$ hour.
\begin{figure}
  \centering
\includegraphics[width= 0.49 \textwidth]{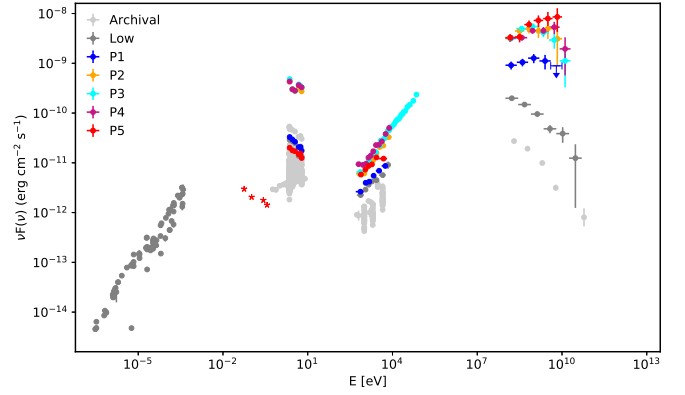}
    \caption{The broadband SEDs of CTA 102 in the selected periods. The archival data are shown in light gray.}%
    \label{Seds}
\end{figure}

\section{Spectral evolution}\label{sec:3}
A $"$Light curve/SED movie$"$ is made for a better understanding of the spectral evolution in different bands. For each adaptively binned interval, using the estimated photon index and flux, the \gray{} spectra are calculated by dividing the (0.16-300) GeV interval into five logarithmically equal bins. These \gray{} spectra are combined with the UV/optical/X-ray (if available) data to make SEDs. As moving from bin to bin, the spectra in all bands can be compared and their evolution in time seen.\\
The movie is uploaded here \href{https://youtu.be/K9WWWSy6W8U}{\nolinkurl{youtu.be/K9WWWSy6W8U}}, where the time period from MJD 57620 to MJD 57950 coinciding with the most active \gray{} emitting state is presented. Up to $\simeq$ MJD 57730, the emission from the source had a soft photon index $\Gamma\geq2.0$ and a maximum flux around $\simeq10^{-10}\:{\rm erg\:cm^{-2}\:s^{-1}}$, which afterwards exceeded $10^{-9}\:{\rm erg\:cm^{-2}\:s^{-1}}$ with hard \gray{} photon indices. Starting from MJD 57765, the flux dropped to its original level and the \gray{} photon index softened. Around MJD 57800, when the flux increased again, the photon indices were $\Gamma\simeq2.0$, implying a flat spectrum of the source in ($\nu-\nu F_{\rm \nu}$) representation. These spectral evolutions once more confirm a harder-when-brighter trend.
\begin{table}[t!]
\scriptsize
 \begin{center}
 \caption{Parameters of spectral analysis}\label{tab:results}
 \begin{tabular}{c c c c }
 \hline
 \hline
  \multicolumn{4}{c}{\fermi{}{}} \\
 \hline
  Period   & Photon Index \tablenotemark{a} & Flux\tablenotemark{b} & $\sigma$\tablenotemark{c}   \\
  \hline
  low  & $2.39\pm0.03$ & $1.13\pm0.04$ & 61.4\\
  P1 & $2.01\pm0.09$ & $6.34\pm0.72$ & 25.0 \\  
  P2 & $1.93\pm0.08$ & $24.17\pm2.43$ & 33.4\\
  P3 &$1.96\pm0.04$ & $24.74\pm1.31$ & 56.5 \\
  P4 &$1.93\pm0.05$ & $21.72\pm1.40$ & 48.7 \\
  P5 &$1.81\pm0.08$ & $25.14\pm2.65$ & 31.0 \\
 \hline
 \multicolumn{4}{c}{Swift-XRT} \\
 \hline
  Period & Photon Index \tablenotemark{d} & Unabsorbed Flux \tablenotemark{e} & $\chi^2_{\rm red}$ (d.o.f.) \\
  \hline
low  &  $1.44\pm0.05$ & $1.45\pm0.07$ & 1.10(39) \\
P1 &  $1.41\pm0.05$ &  $1.91\pm0.09$ & 0.77(52) \\
P2 &  $1.23\pm0.05$ &  $4.79\pm0.22$ & 0.97(53) \\
P3 &  $1.25\pm0.04$ &  $5.75\pm0.13$ & 1.26(84) \\
P4 &  $1.32\pm0.04$ &  $6.46\pm0.15$ & 1.20(75) \\
P5 &  $1.56\pm0.06$ &  $3.31\pm0.15$ & 0.91(31) \\
\hline
 \multicolumn{4}{c}{NuSTAR} \\
 \hline  
 P4\tablenotemark{f}  & $1.32\pm0.005$ & $29.36\pm0.20$ & 0.97(1131)  \\
  \hline
\multicolumn{4}{l}{%
  \begin{minipage}{0.45 \textwidth}
 \tablenotetext{}  Notes:
\tablenotetext{a}{\gray{} photon index from likelihood analysis.}
 \tablenotetext{b}{\gray{} flux in the $0.1-300$ GeV energy range in units of $10^{-7}\:{\rm photon\:cm^{-2}\:s^{-1}}$.}
 \tablenotetext{c}{Detection significance}
 \tablenotetext{d}{X-ray photon index.}
 \tablenotetext{e}{0.3--10 keV X-ray flux corrected for the Galactic absorption in units of $\times$10$^{-11}$ erg cm$^{-2}$ s$^{-1}$.}
 \tablenotetext{f}{X-ray flux and photon index are measured in the energy range 3--79 keV}
\end{minipage}%
}\\
 \end{tabular}
\end{center}

\end{table}
\subsection{Spectral analysis}
The data from the following periods are considered for the spectral analyses:
\begin{itemize}
\item[] 
{\it low state (when the source was not flaring in the \gray{} band)}: when X-ray and \gray{} fluxes were in their average levels: from Swift observations, Obsid: 33509078, 33509079, 33509085, 33509086 and 33509091 were analyzed by merging them to increase the exposure and statistics as they have similar X-ray flux and photon indices while  a few intervals, when the source flux exceeded $9\times10^{-7} \:{\rm photon\:cm^{-2}\:s^{-1}}$, were excluded from the contemporaneously obtained \gray{} data. This period corresponds to the pre-flaring state, allowing to investigate the source emission spectrum before the major flare.

\item[] {\it Period 1 (P1):} MJD 57625.06-57625.39 when the source was in the bright \gray{} state coinciding with XRT observations (Obsid: 33509022 and 33509023, merged during the analyses). 

\item[] {\it Period (P2):} MJD 57738.02-57738.08, bright \gray{} period coinciding with the Swift Obsid: 33509106. 

\item[] {\it Period 3 (P3):} $\simeq3.11$ hour period centered on MJD 57752.52, corresponding to a bright \gray{} state coinciding with Swift (Obsid: 33509112 and 88026001, merged) and NuSTAR observations. 
   
\item[] {\it Period 4 (P4):} $\simeq8.06$ hour period centered on MJD 57759.62, corresponding to the period when the highest X-ray flux was observed (Obsid: 33509115). 

\item[] {\it Period 5 (P5):} $\simeq14.66$ min period centered on MJD 57862.15, corresponding to another peak of \gray{} emission and available quasi-simultaneous Swift observation on the next day (Obsid: 33509121).

\end{itemize}
During the unbinned likelihood analyses of \fermi{} data, the spectrum of CTA 102 has been modeled using a power-law function with the normalization and index as free parameters. Then, the  SEDs are calculated by fixing the power-law index of CTA 102 and running {\it gtlike} separately for smaller energy bins of equal width in log scale. For the spectral analyses the Swift data were binned to have at least 20 counts per bin and then fitted using $\chi^2$ minimization technique. Then, in order to increase the significance of individual points in the SEDs calculations, a denser rebinning was applied, restricting the energy range to $>\;0.5$ keV. The results of analyses (both X-ray and \gray) are given in Table \ref{tab:results} and the corresponding spectra shown in Fig. \ref{Seds}.\\
The \gray{} emission spectra in the low state extended up to $\sim10$ GeV with a soft photon index of $\Gamma=2.39\pm0.03$ while it hardened during the flares, e.g., $\Gamma=1.81\pm0.08$ during P5. There is an indication of deviation of the model with respect to the data above several GeV during P3 (cyan data in Fig. \ref{Seds}). An alternative fit with functions in the form of $dN/dE\sim E_{\gamma}^{-\alpha}\:\times Exp(-E_{\gamma}/E_{cut})$ and $dN/dE\sim (E_{\gamma}/E_{\rm br})^{-(\alpha+\beta log(E_{\gamma}/E_{\rm br}))}$ were applied to check whether the curvature in the spectrum is statistically significant. The first fit resulted in $\alpha=1.64\pm0.09$ and $E_{\rm cut}=3.84\pm1.21$ GeV which is preferred over the simple power-law modeling (comparing log likelihood ratio tests) with a significance of $4.81\:\sigma$. The second fit with $\alpha=1.58\pm0.10$ and $\beta=0.21\pm0.05$ is preferred with a significance of $5.2\:\sigma$. The breaks in the emission spectra can be expected from pair production in BLR \citep{2010ApJ...717L.118P} or can be related with the breaks in the emitting electron spectra \citep{2010ApJ...710.1271A}. The possible origin of the curvature in the GeV spectra should be investigated deeper, with more detailed spectral analyses of single as well as several flaring periods, which is beyond the scope of the current paper.
\section{Broadband SEDs}\label{sec:4}
Fig. \ref{Seds} shows the broadband SEDs of CTA 102 in its low and active periods together with the archival radio-X-ray data (light gray) from ASI science data center. The WISE IR data are highlighted by red asterisk which are most probably due to the torus emission as the recent studies show that the detection rate of almost all \gray{} blazars was high in the WISE all-sky survey \citep{2016ApJ...827...67M}. The comparison shows that during the considered periods the fluxes in the optical/X-ray and \gray{} bands exceed the averaged archival data: the increase is more significant in the optical/UV band. This increase in all bands is expected as the selected periods correspond to the pre-flaring, flaring and post flaring states, and the source shows different emission properties as compared with the averaged spectrum.\\
Comparing our selected period {\it i)} the low-energy component increased while its peak frequency remained relatively constant ($\leq\:10^{15}$ Hz), {\it ii)} the second component increased and shifted to HEs with a strong Compton peak dominance and {\it iii)} the UV/optical, X-ray and \gray{} fluxes contemporaneously increased in P2, P3 and P4, while the emission in the UV/optical and X-ray bands was relatively constant in P1 and P5.\\
The blazar flares can be explained by the changes in the magnetic field, in the emitting region size and its distance from the black hole, bulk Lorentz factor, particle energy distribution, etc. \citep{paggi}. For example, both emission components will be shifted to HEs when the particles are effectively re-accelerated. Only the HE component will increase when the contribution of the external photon fields starts to dominate, for example, due to the changes in the location of the emitting region \citep{paggi}. However, these are not unique 
models for explaining the flaring events. Another possibility is the geometrical interpretation of the origin of flares, the case when the jet regions may have different viewing angles. Such a model with a twisted inhomogeneous jet was already applied to explain the emission from CTA 102 jet in the optical, infrared and radio bands \citep{2017Natur.552..374R}. The photons of different energy come from the jet regions which have different orientations (hence, different Doppler boosting factors) because of the curvature of the jet. \\
The SEDs obtained in the low state, P1 and P5 showing different features, and in the bright P2 have been modeled. In order to account for Compton dominance, we assume the bulk Lorentz factor ($\delta$ which equals to the bulk Lorentz factor for small viewing angles, $\delta\simeq \Gamma$) of the emitting region increased from 10 in the  low  to 20 in the active states (these are typical values estimated for FSRQs \citep{ghistav}). 
When the SEDs in the low state and in P2 are modeled, the emission from a compact region inside and outside the BLR is discussed. Instead, when modeling the periods with lacking correlation in the $\gamma$-ray and UV/optical/X-ray bands, we assume the emission from the radio to X-rays is produced in the extended and slow-moving region unrelated to the flaring component, while the HE \grays{} come from a compact and fast-moving region outside BLR \citep{tavecchio11}.
\subsection{Modeling the SEDs}
The SEDs are fitted within a leptonic scenario that includes synchrotron/Synchrotron Self-Compton (SSC) \citep{ghisellini, bloom, maraschi} and External Inverse-Compton (EIC) \citep{sikora} models. A spherical emission region ("blob") with a radius of $R$ and $B$ magnetic field carries relativistic electrons with a  $N^{\prime}_{\rm e}(E^{\prime}_{\rm e})= N^{\prime}_{0}\:\left( E^{\prime}_{e}/m_{e}\:c^2\right)^{-\alpha}\:Exp[-E^{\prime}_{\rm e}/E^{\prime}_{\rm cut}]$ distribution for $E^{\prime}_{\rm e}\geq E^{\prime}_{\rm min}$ where $E^{\prime}_{\rm min}$ is the minimum electron energy. The size of the emitting region can be inferred from the observed e-folding timescale of $4.08$ hours from the $R\leq\delta\:c\:t/(1+z)\approx\delta\times2.16\times10^{14}$ cm relation.  For the extended emission component, a region with a ten times larger radius ($\simeq4\times10^{16}$ cm) will be used. \\
The low-energy component is modeled by synchrotron emission while for the Inverse Compton (IC) scattering the photons from synchrotron emission, from BLR and dusty torus will be taken into account. The density of BLR ($u_{BLR}$) and dusty torus ($u_{dust}$) are calculated as functions of the distance $r$ from the black hole by the formulae, (e.g., \citet{sikora09})
\begin{equation}
u_{\rm BLR} (r)=\frac{ L_{\rm BLR}}{4\pi r^2_{\rm BLR}c[1+(r/r_{\rm BLR})^3]},\
\label{u1}
\end{equation}
\begin{equation}
u_{\rm dust} (r)=\frac{L_{\rm dust}}{4\pi r^2_{\rm dust}c[1+(r/r_{\rm dust})^4]}.\
\label{u2}
\end{equation} 
The estimated size and luminosity of BLR correspondingly are $r_{\rm BLR}=6.73\times10^{17}$ cm and $L_{\rm BLR}=4.14\times10^{45}\:{\rm erg\:s^{-1}}$ \citep{pian}. The disk luminosity is $L_{\rm disk}=10\times L_{\rm BLR}\simeq 4.14\times10^{46}{\rm erg\:s^{-1}}$ (assuming its 10\% is reprocessed into BLR radiation) then the size and luminosity of torus will be $R_{dust}=10^{18}\:(L_{\rm disc}/10^{45})^{0.5}= 6.43\times10^{18}$ cm \citep{nenkova} and $L_{\it dust}=\eta\:L_{\it disc}=1.24\times10^{46}{\rm erg\:s^{-1}}$ ($\eta=0.6$, \citep{ghisellini2009}) a little larger than the value from tentative detection of dust emission in CTA 102 \citep{malmrose}. Moreover, reproducing the near-IR data presented in Fig. \ref{Seds} with a blackbody component requires a luminosity of a few times $10^{46}{\rm erg\:s^{-1}}$ in agreement with the value used. We adopt an effective temperature $T_{\rm BLR}=10^4\ $K for the BLR radiation and $T=10^{3}$ K for dusty torus.\\
The model free parameters and their uncertainties are estimated using a Markov Chain Monte Carlo (MCMC) method. We have modified the {\it naima} package \citep{zabalza} and the spectral model parameters have been derived through MCMC sampling of their likelihood distributions. For the model free parameters the following expected ranges are considered: $1.5\leq\alpha\leq10$, $0.511\:{\rm MeV}\leq E^\prime_{\rm cut,\:min}\leq10\:{\rm TeV}$, and $N_0$ and $B$ are defined as positive parameters.
\section{Results and Discussion}\label{sec:5}
The broadband emission from CTA 102 during its bright period in 2016-2018 was investigated. In the \gray{} band, during several periods the flux exceeded $10^{-5}\:{\rm photon\:cm^{-2}\:s^{-1}}$ with the maximum being $(3.55\pm0.55)\times10^{-5}\:{\rm photon\:cm^{-2}\:s^{-1}}$ (above 100 MeV) observed on MJD 57738.47 which corresponds to an apparent isotropic \gray{} luminosity of $L_{\gamma}=3.25\times10^{50}\:{\rm erg\:s^{-1}}$ (for a distance of $d_{\rm L}=6.91\:{\rm Gpc}$). This is one of the highest \gray{} luminosities observed from blazars so far (e.g., see \citet{nalewajko}). In the proper frame of the jet, the power emitted in the \gray{} band is $\sim L_{\gamma}/2\delta^2=4.06\times10^{47}\:{\rm erg\:s^{-1}}$ for $\delta=20$ which is higher than $L_{\rm disk}$ in agreement with the results by \citet{ghisellini14}. During this bright period, on a 6-h timescale, the apparent luminosity was $\simeq2.0\times10^{50}\:{\rm erg\:s^{-1}}$ with the rate of change $L/\Delta t\simeq1.89\times10^{46}\:{\rm erg\:s^{-2}}$ (using $\Delta t=6\:{\rm h}/(1+z)\simeq1.06\times10^{4}$ s), slightly higher than that observed from 3C 454.3 \citep{2011ApJ...733L..26A} and well above the Elliot-Shapiro relation \citep{1974ApJ...192L...3E}.\\
The photon index varies as well: the hardest was $1.61\pm 0.10$ observed on MJD 57752.45 which is unusual for FSRQs (having an average photon index of $2.4$ \citep{ackermancatalog}), while on MJD 57528.63 it was as soft as $3.08\pm0.23$. The hardest and softest photon indices were observed during the active and low states, confirming the harder-when-brighter trend. The HE photons ($>10$ GeV) were mostly emitted during the active period of MJD 57700-57800, the highest energy photon being 97.93 GeV. The fractional variability parameter $F_{\rm var}$ shows that the variability is stronger in the \gray{} band ($F_{\rm var}>0.9$), increasing at higher energies.\\
The observed flares are asymmetric which might be due to different relations between particle acceleration and emission timescales. For example, the flares decrease much faster (F1-F3) when the accelerated particles start to escape from the emitting region or the cooling time gradually increases. Whereas, the flare will appear with a fast rise and a slow decay trend (F4) when the fast injected energetic particles loose energy or escape from the regions for a longer time. The observed shortest e-folding time is $\simeq4.1$ hours, inferring that the emitting region is compact. However, during the brightest periods of $\sim$MJD 57738.0 and $\sim$MJD 57752.0, several minutes of observations were already enough to have $> 14.3\sigma$ detection significance, implying shorter time scale variability cannot be excluded (see \citet{shukla} for detailed analysis in shorter periods).\\
Contemporaneous increase in the UV/optical and X-ray bands were also observed during some bright \gray{} periods. In the X-ray band (0.3-10 keV), the maximum flux is $(6.71\pm 0.21)\times10^{-11}\:{\rm erg\:cm^{-2} s^{-1}}$ and the photon index hardens in the bright periods. Comparing the Swift UVOT data obtained in different periods (see Fig. \ref{Seds} and SED/light curve movie) one can see a clear indication of flux increase in the UV/optical bands as well.
\begin{figure*}
  \centering
\includegraphics[width= 0.49 \textwidth]{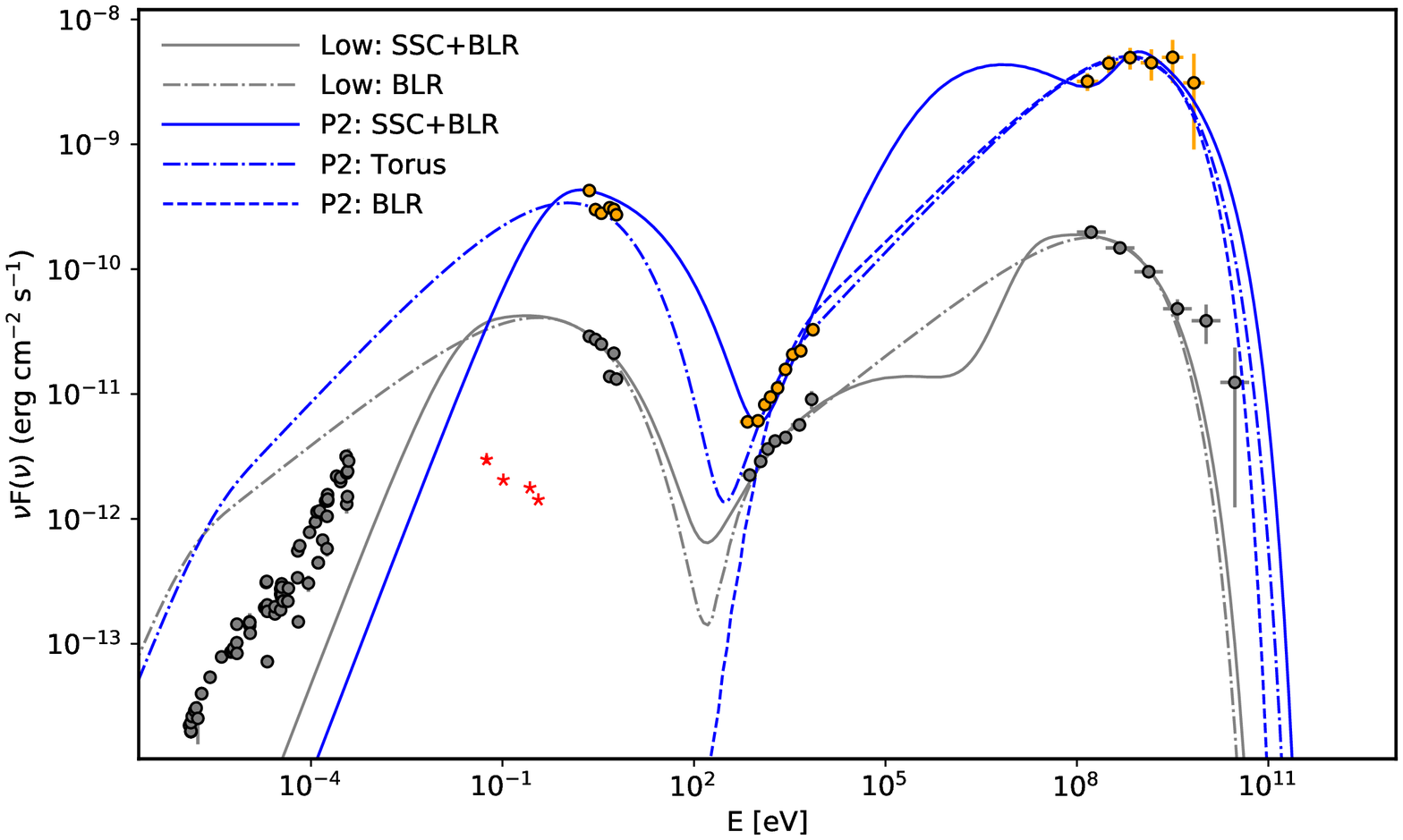}
\includegraphics[width= 0.49 \textwidth]{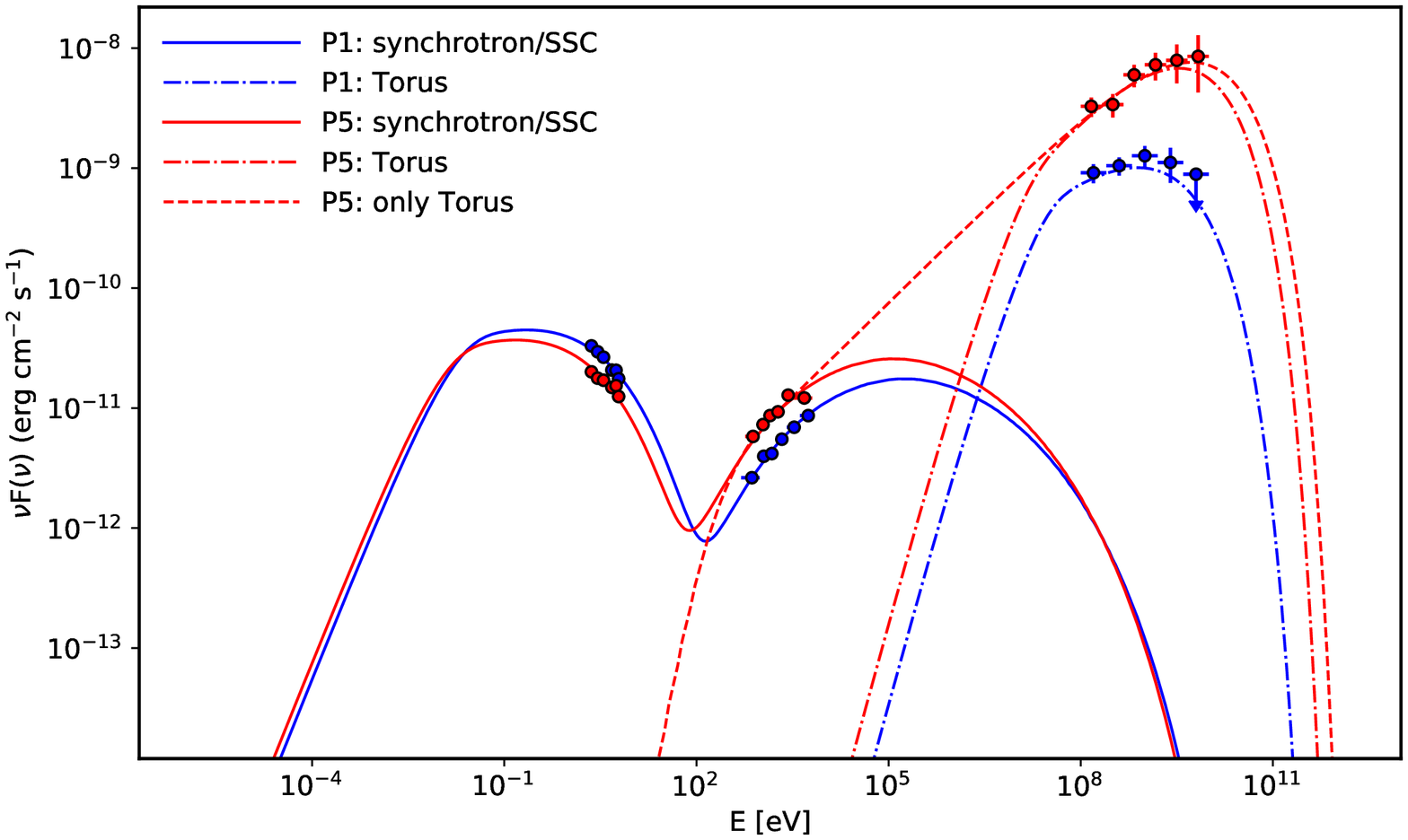}
    \caption{Modeling of the broadband SEDs of CTA 102 during the low state and P2 (left panel, gray and orange, respectively) and P1 and P5 (right panel, blue and red, respectively). The model parameters are given in Table \ref{parres}. For the models applied see the text.}%
    \label{S}
\end{figure*}

\subsection{The origin of the emission}
\begin{deluxetable*}{l|cr|c|cc|cc}
\tabletypesize{\footnotesize}
\tablecaption{Parameters best describing the multiwavelength emission in different periods. \label{parres}}
\tablehead{\colhead{} & \multicolumn{2}{|c|} {low }&\multicolumn{1}{c|}{P1} & \multicolumn{2}{c|}{P2} & \multicolumn{2}{c}{P5} \\\hline
& SSC+BLR  &
BLR         & 
compact &
SSC+BLR  &
Torus &
compact &
Torus 
}
\startdata
$\delta$ & 10 & 10 & 20 & 20 & 20 & 20 & 20\\
$\alpha$ & $2.51\pm0.11$ & $2.19\pm0.02$ &  $2.12\pm0.54$ &  $2.79\pm0.44$ & $1.91\pm0.03$ & $1.78\pm0.52$ & $1.95\pm0.03$  \\
$E_{\rm min}$[MeV] & $68.25\pm5.27$ & $0.54\pm0.03$ &  $155.59\pm109.18$ & $227.25\pm26.43$ & $1.38\pm0.15$ & $121.33\pm67.33$ & $0.63\pm0.09$ \\  
$E_{\rm c}$[GeV] & $0.67\pm0.1$ & $0.49\pm0.04$ &  $1.42\pm0.81$ & $1.32\pm0.43$ & $0.98\pm0.05$ & $2.36\pm1.54$ & $3.85\pm1.57$ \\
$E_{\rm max}$ [TeV] & $0.57\pm0.31$ & $0.49\pm 0.31$ & $0.48 \pm 0.34$ & $0.50\pm0.30$ & $0.41\pm0.18$ & $0.58\pm0.25 $ & $0.54\pm 0.31$\\
$B$[G] & $5.40\pm0.13$ & $5.37\pm0.14$ &  $0.23\pm0.29$ & $6.10\pm0.50$ & $1.01\pm0.003$ &  $0.004\pm0.042$ & $0.015\pm0.049$  \\
$L_{\rm B}[{\rm erg\:s^{-1}}]$ & $1.75\times10^{46}$ & $1.73\times10^{46}$ & $1.47\times10^{42}$ & $1.04\times10^{45}$ & $2.86\times10^{43}$ &  $3.86\times10^{38}$ & $6.44\times10^{39}$\\
$L_{\rm e}[{\rm erg\:s^{-1}}]$ & $4.66\times10^{44}$ & $2.90\times10^{45}$ & $1.73\times10^{46}$ & $2.84\times10^{45}$ & $2.74\times10^{47}$ & $7.33\times10^{46}$ & $1.97\times10^{47}$
\enddata
\end{deluxetable*}
Initially, we modeled the SED observed in the low state (Fig. \ref{S}; left panel). The radio data are treated as upper limits during the modeling, as the emission in this band is produced from the low-energy electrons which are perhaps from much extended regions. We note that the IR flux predicted by the models exceeds the archival IR data $\sim200$ times in the flaring (P2) and $28.7$ times in the selected low states (see Fig. \ref{S}; left panel), implying that the non-thermal synchrotron emission from the jet dominates over the other emission components. When the IC scatterings of both synchrotron and BLR photons are considered, the X-ray data allow to measure $E_{\rm min}^{\prime}=68.25\pm5.27$ MeV and $\alpha=2.51\pm0.11$. In order to explain the observed UV/optical data, a $E_{\rm c}^{\prime}=0.67\pm0.1$ GeV cut-off is required which makes the SSC component to decay in sub-MeV band and the HE data are described only by IC of BLR photons. Alternatively, both X-ray and \gray{} data can be described by IC scattering of BLR photons (dot-dashed gray line in Fig. \ref{S}) but the low-energy tail of IC spectra can reproduce the X-ray data only if $\gamma_{\rm min}=E_{\rm e}/m_{e}c^2$ is close to unity \citep{cellotti}. In this case, however, the synchrotron emission of these low energy electrons with $E_{\rm min}=0.54\pm0.03$ MeV will exceed the observed radio flux, making this scenario unlikely.\\  
\paragraph{P2}
Fig. \ref{S} (left panel) shows the modeling of the SED observed in P2, considering the synchrotron and BLR photons (SSC+BLR, solid line) and then only BLR (dashed line) and only torus (dot-dashed line) photons. When the emitting region is within BLR (SSC+BLR), the hard X-ray spectra $1.23\pm0.05$ can be explained only when $E_{\rm min}^{\prime}=227.25\pm26.43$ MeV and $\alpha=2.79\pm0.44$, while $E_{\rm c}^{\prime}=1.32\pm0.43$ GeV and $B=6.10\pm0.50$ G are estimated from the low-energy component.
Also, the external photon fields can dominate for the IC scattering as their density will increase $\Gamma^2$ times in the jet frame. For example, the required parameters (especially $B$) can be somewhat softened when only the IC of torus photons is considered (see Table \ref{parres}). In the case of only BLR photons, the low-energy tail of IC spectra will decline at $\sim\gamma^{2}\:\epsilon_{\rm BLR}\simeq0.52$ keV (dashed line in Fig. \ref{S} left panel), contradicting Swift XRT data (unless lower $\delta$ is used). This modeling shows that during the bright \gray{} periods the emission can be also produced outside the BLR. At low energies, the model flux overpredicts noncontemporaneous radio data, but when taking the synchrotron self-absorption into account, which dominates below the frequencies $\sim10^{13}$ Hz (calculated following \citet{1979rpa..book.....R}), the synchrotron flux will be below the radio data. We note that simultaneous observations at low energies, which are missing in this case, are crucial for better constraining of the model free parameters and for deriving some limits/constraints on the source emission properties. As the models presented in Fig. \ref{S} (left panel) predict different spectra and fluxes at GHz or mid-IR range, the observations at these bands can be also used to distinguish between these two models.
\paragraph{P1 and P5} Fig. \ref{S} (right panel) shows the results of a two-zone SEDs modeling. For the emission from the extended blob we fixed all the parameters, except $B$ and $N_{0}$, to the values obtained from the fitting of the SED in the low state, as in the UV/optical and X-ray bands the flux and photon indices did not change significantly (Fig. \ref{Seds}). In addition, all the parameters of the compact blob are free, but it is required that its synchrotron emission has no contribution at lower energies.\\
As compared with the low state, the magnetic field in the extended blobs is estimated to be low, $5.05\pm0.08$ G and $3.43\pm0.05$ G for P1 and P5, respectively, implying the modest X-ray flux changes are related with the increase of electron density. The \gray{} emission is produced in the interaction of fresh electrons (hard power law index $\leq2.1$) with the torus photons in the compact, fast-moving and particle-dominated blob $U_{\rm e}/U_{\rm B}\geq10^{4}$ (Fig. \ref{S} right panel). The cut-off energies (defined by the last point in the \fermi{} data) should be considered as lower limits, since there is no indication of break in the \gray{} spectra. In Fig. \ref{S} (right panel) the red dot-dashed line shows an alternative modeling, when both X-ray and \gray{} data are modeled by the IC scattering of torus photons. Within such a scenario, the flare is mainly due to the injection/cooling of $>$ 10 GeV electrons, which are affecting only the HE spectra having small contribution to the X-ray band (e.g., the density at lower energies increases due to the cooling of HE electrons). Again, the low energy component should be necessarily produced in a different blob, otherwise its relatively constant peak frequency cannot be explained. 
\paragraph{Jet energetics}
The total power of the jet, $L_{\rm jet}=L_{\rm B}+L_{\rm e}$ where $L_{B}=\pi c R_b^2 \Gamma^2 U_{B}$ and $L_{e}=\pi c R_b^2 \Gamma^2 U_{e}$ (e.g., \citep{cellot}), is of the order of $L_{\rm jet}\simeq2\times10^{46}\:{\rm erg \:s^{-1}}$ in the low state and can be as large as $\simeq3\times10^{47}\:{\rm erg \:s^{-1}}$ during the flares.\\
When the low and high energy components are contemporaneously increased the required maximum energy of electrons ($E_{\rm c}$) reaches only a few GeV constrained by the low energy data (the energy of synchrotron photons is proportional to $\sim \delta\:B\:E_{\rm e}^2$). Therefore, during these intense \gray{} flares, the acceleration mechanisms are not effective enough or the electrons cool faster and do not reach HEs. On the other hand, when the \gray{} and UV/optical/X-ray fluxes are uncorrelated, the \grays{} are perhaps produced in a different part of the jet that contains fresh electrons which can emit up to HE and VHE bands.
\section{Conclusions}\label{sec:6}
We report the results on the observations of CTA 102 in the UV/optical, X-ray and \gray{} bands from January 2016 to January 2018 when the source was in the bright and active states. Generally, the flares are roughly correlated in all these bands but the variability is more prominent in the \gray{} band with several bright flares when the \gray{} flux is substantially increased and the photon index is hardened, showing a harder-when-brighter trend. The measured hardest photon index $\Gamma=1.61\pm 0.10$ significantly differs from the average \gray{} photon index of CTA 102 and is unusual for FSRQs. The highest \gray{} flux measured by \fermi{} is $(3.55\pm0.55)\times10^{-5}\:{\rm photon\:cm^{-2}\:s^{-1}}$ (above 100 MeV) observed on MJD 57738.47, corresponding to an extremely high isotropic \gray{} luminosity of $L_{\gamma}=3.25\times10^{50}\:{\rm erg\:s^{-1}}$. \\
We discussed the origin of the multiwavelength emission from CTA 102 in the framework of the one-zone and multi-zone synchrotron, SSC and EIC scenarios. We assumed a compact  ($R\leq \delta\times2.16\times10^{14}$ cm inferred from $4.08$ hours \gray{} flux variation) blob inside and outside the BLR. In a single emitting region, the inverse-Compton up-scattering of both synchrotron and BLR photons can explain the data observed in the low state, whereas the contribution of torus photons is essential in the flaring periods. When in the flaring periods the fluxes in the UV/optical, X-ray and \gray{} bands are unrelated, the two-zone models (with an extended blob inside and a compact fast-moving one outside the BLR) can well explain the observed data under reasonable assumptions on the required parameters. These periods appear to be more favorable for the HE emission from CTA 102 as the emitting electrons have higher cut-off energies and harder power-law indicies. Most likely, the emission in these periods is produced in the regions outside BLR that contain fresh electrons which dominantly cool due to the inverse-Compton scattering making the variability more evident in the \gray{} band.
\section*{acknowledgements}
 We thank the anonymous referee for constructive comments that improved the paper.
\bibliography{references}
\bibliographystyle{aasjournal}

\end{document}